\documentclass{aa}  
\usepackage{graphicx}
\usepackage{txfonts}
\begin{document} 

\title{The dependency of AGN infrared color-selection on source luminosity and obscuration}
\subtitle{An observational perspective in CDFS and COSMOS}

\author{H. Messias\inst{1,2} \and J. Afonso\inst{1} \and M. Salvato\inst{3,4} \and B. Mobasher\inst{5} \and A. M. Hopkins\inst{6}}

\institute{
Centro de Astronomia e Astrof\'{i}sica da Universidade de Lisboa, Observat\'orio Astron\'omico de Lisboa, Tapada da Ajuda, 1349-018 Lisboa, Portugal.\\
\email{hmessias@oal.ul.pt}
\and
Departamento de Astronom\'ia, Av. Esteban Iturra 6to piso, Facultad de Ciencias F\'isicas y Matem\'aticas, Universidad de Concepci\'on, Chile.
\and
Max Planck for Extraterrestrial Physics, Giessembachstrasse 1, 85748, Garching, Germany
\and
Excellence Cluster, Boltzmann Strasse 2, 85748, Garching, Germany
\and
University of California, 900 University Ave., Riverside, CA 92521, USA.
\and
Australian Astronomical Observatory, P.O. Box 915, North Ryde, NSW 1670, Australia.}


 
  \abstract
   {}
   {This work addresses the AGN IR-selection dependency on intrinsic source luminosity and obscuration, in order to identify and charaterise biases which could affect conclusions in studies.}
   {We study IR-selected AGN in the \emph{Chandra} Deep Field South (CDFS) survey and in the Cosmological Survey (COSMOS). The AGN sample is divided into low and high X-ray luminosity classes, and into unobscured (type-1) and obscured (type-2) classes by means of X-ray and optical spectroscopy data. Specifically in the X-rays regime, we adopt the intrinsic luminosity taking into account the estimated column density ($\rm{N_H}$). We also take the chance to highlight important differences resulting from adopting different methods to assess AGN obscuration.}
   {In agreement with previous studies, we also find that AGN IR-selection efficiency shows a decrease with decreasing source AGN X-ray luminosity. For the intermediate-luminosity AGN population ($43.3\lesssim\log(\rm{L_X[erg\,s^{-1}]})\lesssim44$), the efficiency also worsens with increasing obscuration ($\rm{N_H}$). The same sample also shows an evolution with cosmic time of the obscured fraction at the highest X-ray luminosities, independently of adopted type-1/type-2 classification method.}
   {We confirm that AGN IR-selection is genuinely biased toward unobscured AGN, but only at intermediate luminosities. At the highest luminosities, where AGN IR-selection is more efficient, there is no obscuration bias. We show that type-1 AGN are intrinsically more luminous than type-2 AGN only at $z\lesssim1.6$, thus resulting in more type-1 AGN being selected the shallower the IR survey is. Based on this study and others in the literature, we conclude that deep hard-X-ray coverages, high-resolution IR imaging, or a combination of IR and radio data are required to recover the lower-luminosity obscured AGN population. In addition wide IR surveys are necessary to recover the rare powerful obscured AGN population. Finally, when the \emph{James Webb Space Telescope} comes online, the broad-band filters 2.0$\,\mu$m, 4.4$\,\mu$m, 7.7$\,\mu$m, and 18$\,\mu$m will be key to disentangle AGN from non-AGN dominated SEDs at depths where spectroscopy becomes impractical.}

   \keywords{galaxies: active --
                galaxies: nuclei --
                galaxies: evolution --
                galaxies: statistics --
                infrared: galaxies --
                X-rays: galaxies
               }

   \maketitle
%

\section{Introduction}

It is believed today that there is a significant fraction of obscured Active Galactic Nuclei (AGN), which still remain unidentified at high-energy spectral-bands even in the current deepest surveys \citep[the strongest evidences coming from X-rays studies, e.g.,][]{Comastri01, Ueda03, Gilli04, Worsley04, Worsley05, TreisterUrry05, Martinez05, Draper09, Burlon11, Ajello12, Moretti12}. Without an uncertain demography of actively accreting super-massive black holes, current galaxy evolution models accounting for AGN feedback \citep[e.g.,][]{Granato04, Springel05, Croton06, Hopkins06, Bower06, Somerville08} likely remain poorly constrained. As a result, alternative means to find such objects, adopting different spectral regimes, ought to be developed.

If large dust column densities are the driver for such extreme obscuration of direct light from the AGN accretion region (instead of just obscuring circumnuclear gas clouds), then infrared (IR) wavelengths, at which the obscuring dust emits \citep[e.g.,][]{Nenkova08,HonigKishimoto10}, are an obvious choice in the search for such extreme objects. Hints for AGN activity at IR wavelengths can be revealed through spectroscopy diagnostics \citep{Laurent00,Armus07,Veilleux09} or photometric techniques. The latter encompass the selection of AGN-like sources by means of\footnote{For a more in depth discussion on the different diagnostics, please consult \citet{Donley08} and \citet{Messias12}.}: color-color constraints \citep{deGrijp85, Miley85, Lacy04, Lacy07, Hatziminaoglou05, Hatziminaoglou10, Stern05, Assef10, Assef13, Hony11, Jarrett11, Donley12,Mateos12, Messias12}; red power-law spectral energy distribution (SED) fitting \citep{Alonso06, Donley07}; excess IR emission unmatched by pure star-forming SEDs \citep{Daddi07}; or large optical-to-IR flux ratios ($f_{24}/f_R>1000$) in addition to bright IR flux cuts and/or extremely red optical-to-near-IR color cuts \citep{Dey08, Fiore08, Polletta08, Donley10}. Finally, the IR spectral regime can be combined with other wave-bands in order to improve AGN selection efficiency \citep[e.g.,][]{Martinez05,Donley05,Richards09,Edelson12}.

For AGN IR-selection to be possible, the circumnuclear dust thermal emission has to dominate the IR SED, thus revealing the features targeted by the techniques mentioned above. Such conditions, however, are frequently not found when studying galaxies hosting low-luminosity AGN, where the host galaxy light dominates instead \citep{Treister06, Cardamone08, Donley08, Eckart10, Petric11, Donley12}. High resolution imaging can probe the very center of a system and avoid this problem \citep[e.g.,][]{Honig10,Asmus11}, but it is only applicable to galaxies in the very local Universe. Also, there are evidences pointing that IR-selection is biased toward unobscured AGN \citep{Miley85, Stern05, Donley07, Cardamone08, Eckart10, Mateos13}. This comes as a shortcoming, given that the main goal of IR-selection is to recover the obscured AGN population missed by even the deepest high-energy surveys. On top of that, by missing low-luminosity AGN \citep[e.g.,][]{Stern05, Eckart10, Mateos13}, IR-selection misses an evolution phase where AGN are believed to spend most of their life span \citep[e.g.,][]{Hopkins05, Hopkins06, Fu10}. Although currently in a low-luminosity phase, these AGN are believed to have induced strong feedback effect in the past in a quasar phase \citep[e.g.,][]{Lipari05,LipariTerlevich06,Hopkins08,Maiolino12} and/or to do so at later times \citep[for instance, induced by galaxy merger events, e.g.,][]{Hopkins08,Lipari09}. Hence, it is important to understand what are the limitations of each of the AGN selection techniques used at different spectral regimes. Such exercise would provide a better knowledge of the incompleteness affecting current AGN demography studies.

In this work, we focus on the AGN IR-selection dependency on source intrinsic luminosity and obscuration by adopting an observational point of view.

We present the sample adopted in this work in Sect.~\ref{sec:samp}. In Sect.~\ref{sec:kikim} we resume the results obtained in \citet[][Paper~I henceforth]{Messias12}. The dependency of the IR selection criteria efficiency on luminosity and obscuration is addressed in Sect.~\ref{sec:typesel}. We present a discussion on the different effects responsible for the completeness of IR criteria in Sect.~\ref{sec:ircomp}. We propose different strategies on the application of IR-selection criteria depending on sample characteristics in Sect.~\ref{sec:strat}. In Sect.~\ref{sec:jwst}, we propose a new criterion based on the filter-set of \emph{JWST}, different from the criterion presented in Paper~I. The conclusions are presented in Sect.~\ref{sec:conc}.

Throughout this paper we use the AB magnitude system\footnote{When necessary the following relations are used:
(K, [3.6], [4.5], [5.8], [8.0])$_{AB}$ = (K, [3.6], [4.5], [5.8], [8.0])$_{Vega}$ + (1.841, 2.79, 3.26, 3.73, 4.40) \citep[][and http://spider.ipac.caltech.edu/staff/gillian/cal.html]{Roche03}.}, assuming a $\Lambda$CDM model with H$_{0} = 70$ km s$^{-1}$ Mpc$^{-1}$, $\Omega_{M} = 0.3$, $\Omega_{\Lambda} = 0.7$. X-ray luminosity ($\rm{L_{XR}}$) corresponds to the 0.5--10\,keV energy range.

\section{The sample} \label{sec:samp}

The sample used in this work has already been described in detail in Sect.~4.1 of Paper~I. Briefly, for the purpose of AGN identification and obscuration assessment, we focused on a galaxy sample with available deep X-ray or optical-spectroscopy data in the \emph{Chandra} Deep Field South \citep[CDFS,][]{Giacconi02} and in the Cosmological Survey \citep[COSMOS,][]{Scoville07}. Only sources with reliable photometry (magnitude error below 0.36, i.e., a flux error smaller than a third of the source flux) in the $K_s$ and IRAC bands are considered. X-ray data come from \citet[][the 4Ms exposure, instead of the 2Ms exposure presented in \citealt{Luo08} adopted in Paper I]{Xue11} in CDFS and \citet{Brusa10} in COSMOS. The same works provide the optical counterparts, whose IR photometry was extracted from the MUSIC \citep[][in CDFS]{Santini09} and \citet[][in COSMOS, see also \citealt{McCracken10}]{Ilbert09} catalogues.  Optical spectroscopy in CDFS has been compiled by \citet{Santini09} in the MUSIC catalogue with extra spectroscopy data analysis coming from \citet{Silverman10}, while the 10k zCOSMOS-bright catalogue \citep{Lilly09} has been considered in the COSMOS field. Detailed photometric redshifts estimates for those X-rays sources with no available spectroscopy are compiled by \citet{Xue11} in CDFS and provided by \citet{Salvato11} in COSMOS.

X-ray and spectroscopic data are considered in order to identify AGN and separate this population into unobscured (type-1) and obscured (type-2) objects. The spectroscopic classification comes from \citet{Santini09} and \citet{Silverman10} in CDFS, and \citet{Lilly09} and \citet{Bongiorno10} in COSMOS. In the X-ray, we follow the criteria proposed by \citet{Szokoly04} where an obscured AGN is considered to have a 0.5--10\,keV luminosity $\log({\rm L_{XR}[erg\,s^{-1}]})>41$, while an unobscured AGN is considered to have $\log({\rm L_{XR}[erg\,s^{-1}]})>42$. The X-ray obscuration classification is based on the column density ($\rm{N_H}$) estimate, where sources with $\log(\rm{N_H[{\rm cm}^{-2}]})\leq22$ are considered unobscured ($\rm{A_1}$), and obscured when $\log(\rm{N_H[{\rm cm}^{-2}]})>22$ ($\rm{A_2}$). However, by estimating $\rm{N_H}$ via X-ray flux ratios (see Paper I), we are inducing a bias toward type-2 objects at the highest redshifts \citep[][and Paper I]{Akylas06,Donley12}. We have thus applied a statistical correction to account for this bias. The correction method is described in detail in the Appendix of Paper I. Its performance is shown for CDFS and COSMOS independently and compared against the Hardness-Ratio (HR, a common alternative) in Appendix~\ref{sec:t12dem} of this paper. The X-ray classification was considered over the optical when the two disagree.

\begin{figure}
\centering
\includegraphics[width=\hsize]{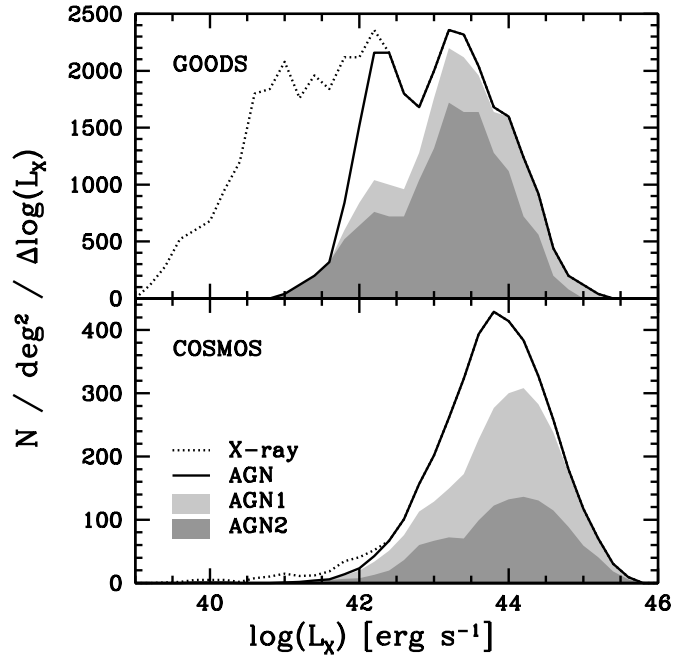}
\caption{The source density distribution with intrinsic X-ray luminosity distribution for CDFS (upper panel, $\sim$140\,arcmin$^2$) and COSMOS (lower panel, $\rm{1.8\,deg^2}$) samples (note the y-axis are different). The distributions were obtained with a moving bin of width $\log(\rm{L_X})=0.6$, with measurements taken each $\log(\rm{L_X})=0.2$. The overall X-ray population is represented by the dotted line, the AGN by the solid line. The AGN population is further separated into the type-1 (light shaded region, $\rm{\log(N_H[cm^{-2}])\leq22}$), type-2 (dark shaded region, $\rm{\log(N_H[cm^{-2}])>22}$), and unknown-type (white region) sub-populations. The regions do not overlap. The unknown-type AGN are those for which data was not enough to allow for an obscuration determination.}
\label{fig:lxdist}
\end{figure}

\begin{figure}
\centering
\includegraphics[width=\hsize]{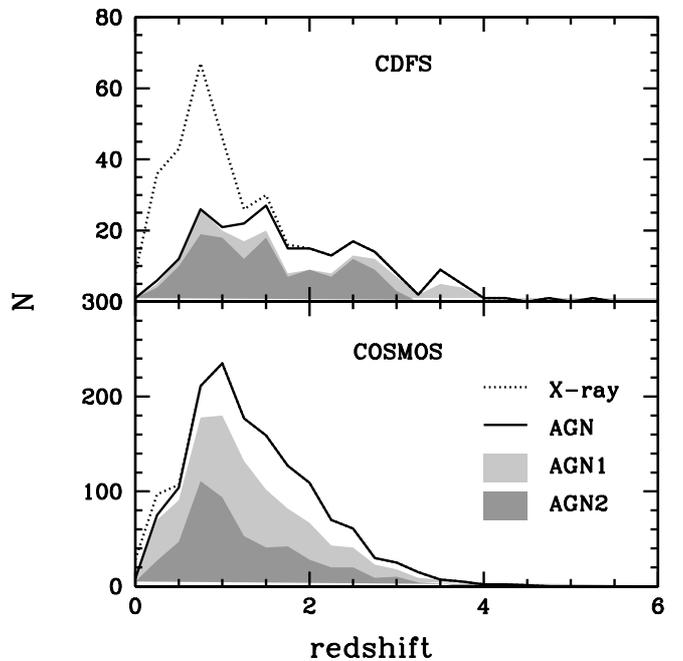}
\caption{The redshift distribution for CDFS (upper panel) and COSMOS (lower panel) samples (note the y-axis are different). The histograms are cumulative, and no moving bin was applied. Line and region coding as in Fig.~\ref{fig:lxdist}.}
\label{fig:redist}
\end{figure}

Table~\ref{tab:samp} details the number of AGN considered in each field, and how they separate into unobscured ($\rm{A_1}$) and obscured ($\rm{A_2}$) AGN, and into high-luminosity ($\rm{A_H}$, $\log({\rm L_{XR}[erg\,s^{-1}]})\geq43.5$\footnote{The boundary $\log({\rm L_{XR}[erg\,s^{-1}]})=43.5$ was chosen so that both populations have a reasonable statistical sample.}) and low-luminosity ($\rm{A_L}$, $\log({\rm L_{XR}[erg\,s^{-1}]})<43.5$\footnote{The lower luminosity-limit depends on the obscuration nature of the source, as described in the previous paragraph.}) AGN. Although upper/lower-limits are considered, about 20\% (in CDFS) and 26\% (in COSMOS) of the AGN do not have an obscuration estimate mostly due to the soft-band (0.5--2\,keV) being $\sim6$ times more sensitive than the hard-band (2--8\,keV). The intrinsic (obscuration corrected) X-ray luminosity and redshift distributions are shown for CDFS and COSMOS in Fig.~\ref{fig:lxdist} and \ref{fig:redist}, respectively, for the X-ray AGN sample, highlighting the type-1 and type-2 AGN components.

\begin{table}
\caption{The sample.}
\label{tab:samp}
\centering
\begin{tabular}{crrrrr}
\hline\hline
Field & $\rm{N_{AGN}}$ & \multicolumn{2}{c}{Obscuration\tablefootmark{a}} & \multicolumn{2}{c}{Luminosity\tablefootmark{b}} \\
& & $\rm{A_1}$ & $\rm{A_2}$ & $\rm{A_H}$ & $\rm{A_L}$ \\
\hline
\multicolumn{6}{c}{[$K_s+\rm{IRAC}$]} \\
CDFS & 218 & 44 & 129 & 69 & 149 \\
COSMOS & 1423 & 543 & 515 & 961 & 457 \\
\hline
\multicolumn{6}{c}{[$K_s+\rm{IRAC+MIPS_{24\,\mu m}}$]} \\
CDFS & 171 & 35 & 106 & 61 & 110 \\
COSMOS & 842 & 381 & 322 & 600 & 238 \\
\hline
\end{tabular}
\tablefoot{
While in the upper group of rows reliable photometry is required --- a magnitude error below 0.36 --- in $K$+IRAC bands, in the lower group of rows we also require reliable 24\,${\mu}m$ photometry.\\
\tablefoottext{a}{The sample is separated into unobscured ($\rm{A_1}$, $\log(\rm{N_H[{\rm cm}^{-2}]})\leq22$ or spectroscopy information) and obscured ($\rm{A_2}$, $\log(\rm{N_H[{\rm cm}^{-2}]})>22$ or spectroscopy information) AGN.}\\
\tablefoottext{b}{The sample is separated into high-luminosity ($\rm{A_H}$, $\log({\rm L_{XR}}[erg\,s^{-1}])\geq43.5$) and low-luminosity ($\rm{A_L}$, $\log({\rm L_{XR}}[erg\,s^{-1}])<43.5$) AGN.}
}
\end{table}

\section{AGN IR-selection criteria} \label{sec:kikim}

In Paper I, we have presented an efficiency analysis of different color-color diagnostics based on X-rays, optical, mid-IR (MIR), millimetre and radio samples. Three of the tested AGN selection criteria, proposed in the literature, consider purely IRAC photometry \citep[][referred henceforth as L07, S05, and D12, respectively]{Lacy04,Lacy07,Stern05,Donley12}, while the two new criteria proposed in Paper I consider $K_s$+IRAC (KI) and $K_s$+IRAC+MIPS$_{24\,\mu\rm{m}}$ (KIM) photometry. The color-color constraints for KI and KIM are as follows:
\begin{equation}
  \rm{KI} \equiv 
  \begin{cases}
    K_s-\left[4.5\right]>0 \\
    \left[4.5\right]-\left[8.0\right]>0
  \end{cases}
\end{equation}
\begin{equation}
  \rm{KIM} \equiv 
  \begin{cases}
    K_s-\left[4.5\right]>0 \\
    \left[8.0\right]-\left[24\right] > 0.5 \\
    \left[8.0\right]-\left[24\right] > -2.9\times(\left[4.5\right]-\left[8.0\right])+2.8
  \end{cases}
\end{equation}

These two criteria were developed with the goal to achieve maximum efficiency (select more AGN sources with less contamination by non-AGN sources) using a minimal number of filters in order to optimise telescope observing time. They were especially designed for deep fields and took into account the \emph{James Webb Space Telescope} (\emph{JWST}) wavelength coverage. The considered filters --- $K_s$, 4.5$\,\mu$m, 8.0$\,\mu$m, and 24$\,\mu$m (or 2.0$\,\mu$m, 4.4$\,\mu$m, 7.7$\,\mu$m, and 18$\,\mu$m during \emph{JWST} science mission, see Sect.~\ref{sec:jwst}) --- are believed to be key to tackle the photometric degeneracy between AGN sources and high-redshift dusty starbursts.

In Paper I, we show that, overall, KI recovers $\sim$50--60\% of the total AGN sample (its completeness, $\mathcal{C}$), and $\sim$50--90\% of the KI-selected sample reveals AGN activity at X-rays and/or optical spectral regimes (its reliability, $\mathcal{R}$). KIM is less complete then KI, recovering $\sim$30--40\% of the total AGN sample, but equally (or more) reliable, with $\sim$75--90\% of the KIM-selected sample revealing AGN activity. The $\mathcal{R}$ values should be regarded as lower limits, given that the X-rays and optical spectral regimes themselves do not provide a complete sample of AGN sources. This also means that $\mathcal{C}$ values should be used for comparison between criteria, and not assumed as real values.

Compared to the \citet[][L07]{Lacy07}, \citet[][S05]{Stern05}, and \citet[][D12]{Donley12} criteria, the most significant improvement by KI and KIM is achieved at $1<z\leq2.5$. At these redshifts, KI (with $\mathcal{C}=40\pm8\%$ and $\mathcal{R}=53\pm12\%$\footnote{Indicated errors refer to Poisson statistical uncertainty.}) and KIM (with $\mathcal{C}=26\pm7\%$ and $\mathcal{R}=64\pm20\%$) are more reliable than L07 ($\mathcal{R}=29\pm5\%$) and S05 ($\mathcal{R}=27\pm6\%$), and KI is more complete than D12 ($\mathcal{C}=19\pm5\%$). At $z\lesssim1$, although our sample did not allow for a conclusive answer, a larger yet shallower WISE sample (likely dominated by $z\lesssim1$ sources) compiled by \citet{Assef13} has distinguished KI and KIM among the best criteria. At higher redshifts ($z\gtrsim2.5$, where criteria using $<8\,\mu$m wave-bands fail), KIM is believed to be the best criteria purely based on a model analysis.

Overall, L07 is the most complete criteria ($\mathcal{C}\sim70-80\%$), but also the least reliable ($\mathcal{R}\sim30-55\%$), while, on the other extreme, D12 is always the most reliable ($\mathcal{R}\sim75-100\%$, although affected by low statistical significance), but the least complete ($\mathcal{C}\sim25-40\%$). S05 selects comparable number of AGN ($\mathcal{C}\sim45-65\%$) as KI, but with a smaller reliability ($\mathcal{C}\sim40-85\%$). KI and KIM are thus believed to be the best alternative, with KI being more complete, yet contaminated by $z\gtrsim2.5$ non-AGN sources. This effect is expected not to affect KIM, making this criterion the best alternative in deep surveys, unless a flux cut is adopted.

\section{Selection of type-1/2 and low-/high-luminosity sources} \label{sec:typesel}

The next step to evaluate the AGN IR-criteria is characterising the selected AGN population in terms of luminosity and obscuration. Previous studies have claimed that IR colour-colour criteria are biased toward unobscured systems \citep[broad-line AGN or type-1 AGN;][]{Stern05, Donley07, Cardamone08, Eckart10, Mateos13}, and tend to select the most luminous objects, missing many low-luminosity ones \citep{Treister06, Cardamone08, Donley08, Eckart10, Donley12}. These tendencies are also assessed in this section.

For the purpose of estimating AGN obscuration, there are many alternatives in the literature, each with its own bias(es). As a result, results will eventually be different in each literature work. In the Appendix~\ref{sec:t12dem} we discuss the bias affecting different approaches depending on survey characteristics and redshift. The analysis presented in Appendix~\ref{sec:t12dem} and in Paper~I, and the limitations of the HR method \citet{Eckart06,Messias10,Brightman12} justify our choice to adopt a statistically corrected $\rm{N_H}$ estimate (described in Paper~I) to assess X-ray obscuration. We believe this is the most reliable approach allowing for such a large AGN sample throughout a wide redshift range.

Given the occasional small number statistics we will be dealing with in this section, we estimate the statistical error as $\pm0.5+\sqrt{\rm{N}+0.25}$, where the negative sign refers to the lower error, and the positive sign to the upper error\footnote{This alternative is proposed by the Collider Detector at Fermilab Statistics Committee as described in: http://www-cdf.fnal.gov/physics/statistics/notes/pois\_eb.txt}. Some advantages of this alternative over the traditional $\sqrt{\rm{N}}$ is that the upper error does not tend to zero for $\rm{N\to0}$, and it retrieves a non-zero lower error for $\rm{N=1}$. For large N, both errors tend to the traditional assumption, $\sqrt{\rm{N}}$.

\subsection{AGN IR-selection dependency on luminosity}\label{sec:selecLum}

\begin{figure}
\centering
\includegraphics[scale=0.36,angle=-90]{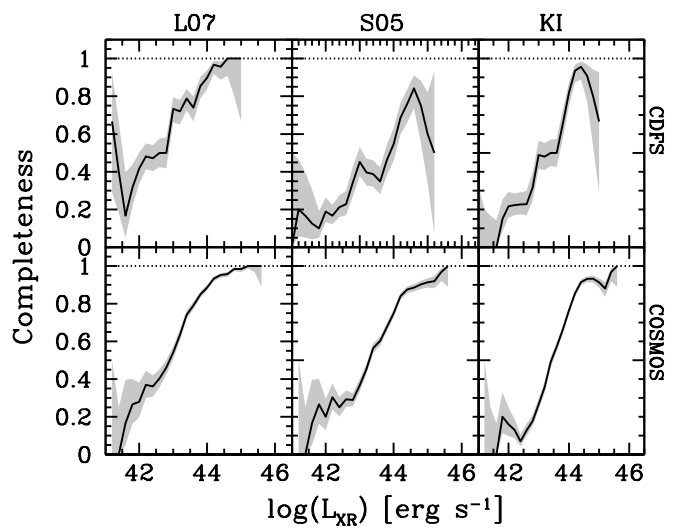}
\caption{The AGN completeness for L07, S05, and KI criteria depending on source X-ray luminosity. The light-grey regions show the associated statistical error. A moving bin is used as described in Fig.~\ref{fig:lxdist}.}
\label{fig:s21c}
\end{figure}

We again emphasize that the aim of IR color-color criteria is the selection of galaxies with an IR SED dominated by AGN light. However, low-luminosity AGN will often not dominate the IR emission of a galaxy, making their IR selection unlikely. This is clearly seen in Fig.~\ref{fig:s21c}, where the completeness of the L07, S05, and KI AGN selection criteria increases significantly with source luminosity, in agreement with previous work \citep{Treister06,Cardamone08,Donley08,Eckart10}. If, as indicated by previous work, type-1 AGN tend to be more luminous than type-2 AGN (see the discussions in \citealt{Treister09}, \citealt{Bongiorno10}, \citealt{Burlon11}, \citealt{Assef13} and references therein), this will then lead to a higher fraction of type-1 objects among IR-selected AGN samples. However, it is important to stress this would not necessarily imply a higher sensitivity of IR criteria toward type-1 AGN, but mostly a natural result of luminosity dependence.

\subsection{Obscuration dependency on luminosity and its evolution with redshift} \label{sec:z12}

\begin{figure*}
\centering
\includegraphics[scale=0.5,angle=-90]{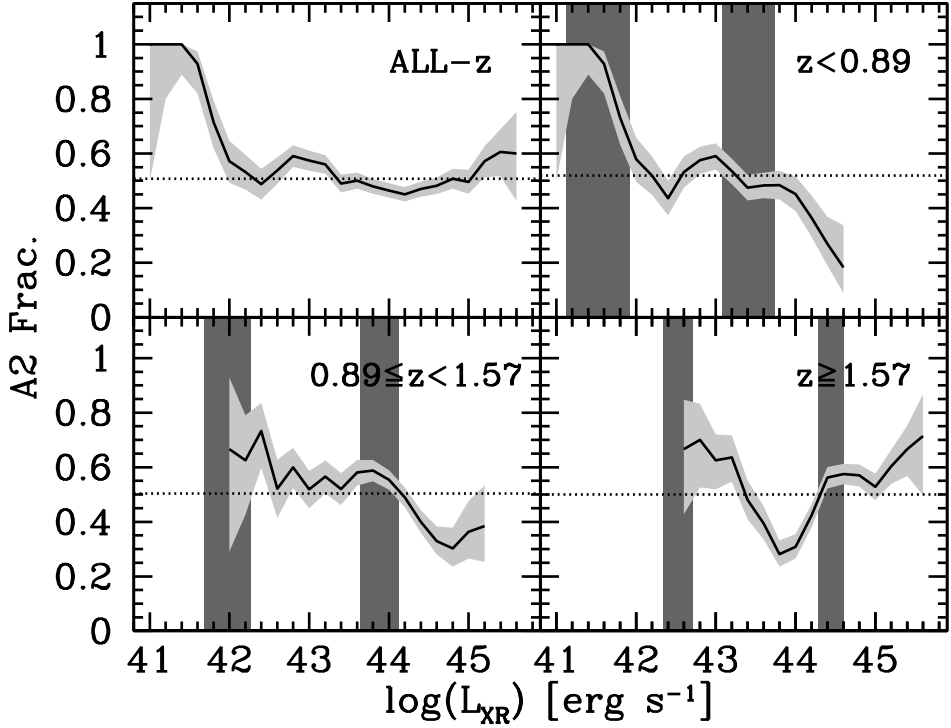}
\caption{The variation of obscured fraction with source X-ray luminosity in CDFS and COSMOS. The ratio is $\rm{A_2/(A_1+A_2)}$, where $\rm{A_1}$ and $\rm{A_2}$ are the numbers of type-1 and type-2 sources, respectively. Here, the two samples are considered together as one. The top left panel considers the whole AGN sample, while the other panels consider, respectively, $z<0.89$, $0.89\leq z<1.57$, and $z\geq1.57$ AGN. A moving bin is used as described in Fig.~\ref{fig:lxdist}. Light-grey shaded regions show the associated statistical error, while dark-grey shaded regions show tentative completeness levels for the CDFS (at low luminosities) and COSMOS (at high luminosities) samples accounting for column densities of up to $\rm{\log(N_H[cm^{-2}])}=24$ (see text). The horizontal dotted line shows the overall $\rm{A_2}$ fraction in each case, $\rm{A_2^{TOT} / ( A_1^{TOT} + A_2^{TOT} )}$.}
\label{fig:s21lumx}
\end{figure*}

As just discussed, a possible natural luminosity-obscuration relation may influence the conclusions. In addition, the type-2 fraction has been reported to increase with redshift \citep[][but see the analysis by \citealt{Lusso13} even though limited to type-1 objects]{Hasinger08,Treister09,Bongiorno10}. Here, too, we briefly address these results. Figure~\ref{fig:s21lumx} shows the variation of type-2 fraction with intrinsic luminosity, where the fraction presented is $\rm{A_2/(A_1+A_2)}$, with $\rm{A_1}$ and $\rm{A_2}$ being the numbers of type-1 and type-2 sources, respectively. For improved display of the results, the CDFS and COSMOS samples have been considered together as one (see Sec.~\ref{sec:t12dem} for their separate behaviour). While the top left panel considers the whole AGN sample, the remaining panels consider three redshift intervals with an equal number of AGN: $z<0.89$ (top right), $0.89\leq z<1.57$ (bottom left), and $z\geq1.57$ (bottom right). The dark-grey shaded regions indicate a crude estimate of the sample completeness for the CDFS and COSMOS X-ray coverage. These are estimated by assuming the intrinsic luminosity of a source at $z=0.89$, 1.57, and 3 yielding an observed flux equal to the survey sensitivity. The luminosity range of the regions reflects the column density range $0\leq\rm{\log(N_H[cm^{-2}])}\leq24$.

Overall, we observe a more-or-less constant type-2 fraction with luminosity at $\log({\rm L_{XR}[erg\,s^{-1}]})\gtrsim42$ of $\sim0.5$. However, when separating the sample into redshift bins, one observes that luminous AGN ($\log({\rm L_{XR}[erg\,s^{-1}]})\gtrsim44$) are mostly unobscured up to $\sim1.6$, while at higher redshifts the opposite is observed. A positive increase of the type-2 fraction at these luminosities is also observed if the HR is used instead of ${\rm N_H}$ to assess unobscured/obscured types (see Sec.~\ref{sec:t12dem}) and is in agreement with the literature. Due to sample incompleteness, we do not attempt to measure how the obscuration-luminosity relation evolves with redshift down to lower luminosities. However, based on the evolution at the highest luminosities, we know it has to change, either in its slope and/or its scaling factor.

\subsection{AGN IR-selection dependency on obscuration}

Now, assuming the efficiency of IR criteria is clearly dependent on source luminosity, as observed in Fig.~\ref{fig:s21c}, the dependency of the type-2 fraction on luminosity (Fig.~\ref{fig:s21lumx}) should dictate the type-2 fraction observed in an AGN sample selected via any AGN IR-criteria. This is, the IR criteria should present the same type-2 fraction observed in the highest luminosity bins of a given sample. In our case, based on Fig.~\ref{fig:s21lumx}, that would mean a fraction of $\sim0.5$ for IR criteria, but less than 0.5 if a shallower sample (limited to low-redshift sources) would be adopted. Does our sample confirm this pure-luminosity dependence or does it imply nevertheless a bias toward type-1 sources, as defended in the literature?

\begin{figure*}
\centering
\includegraphics[width=\hsize]{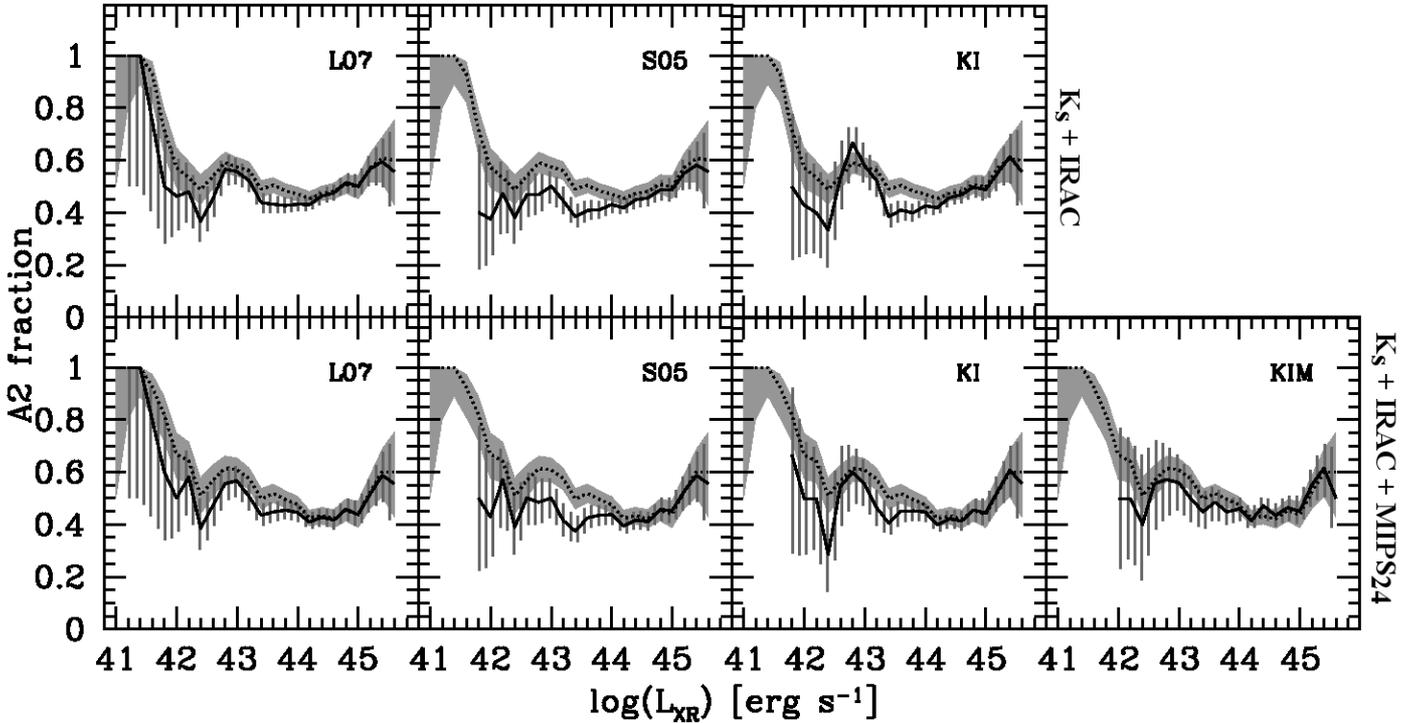}
\caption{The variation of the obscured fraction ($\rm{A_2/(A_1+A_2)}$) with source intrinsic X-ray luminosity (solid line) for L07 (left), S05 (middle left), KI (middle right), and KIM (right). The top row considers sources with reliable photometry in the $K_s+\rm{IRAC}$ bands, while the bottom row considers sources with reliable photometry in the $K_s+\rm{IRAC+MIPS_{24}}$ bands. The CDFS and COSMOS samples are considered together. The shaded and hatched regions show the associated statistical uncertainty. The dotted line refers to the obscured fraction observed in the whole sample (Fig.~\ref{fig:s21lumx}). A moving bin is used as described in Fig.~\ref{fig:lxdist}.}
\label{fig:s21f}
\end{figure*}

Figure~\ref{fig:s21f} helps clarify this point. We recall that we consider the corrected N$_{\rm H}$ and $\rm{L_X^{INT}}$ values as adopted in Paper I. That assumption originates the trend observed in the top left panel in Fig.~\ref{fig:s21lumx} and repeated in Fig.~\ref{fig:s21f} (dotted-line and grey region) for reference. In Fig.~\ref{fig:s21f}, one observes at least S05 and KI criteria being biased toward type-1 AGN at intermediate luminosities ($43.3\lesssim\log(\rm{L_X[erg\,s^{-1}]})\lesssim44$), given that the type-2 fraction  for S05- and KI-selected samples is lower than that observed for the whole sample in this luminosity range. At the highest luminosities the results are consistent with both types being selected with equal efficiency, as the type-2 fraction is unchanged. Limiting the study to sources also with reliable photometry in the $\rm{MIPS_{24}}$ band, the S05 criterion still shows a significant bias toward type-1 AGN.

This is in agreement with the findings of \citet{Treister09}, who noticed a lack of IR excess emission in intermediate luminosity obscured AGN, even though their analysis is mainly spectroscopically based (which could induce a disagreement with our type-1/type-2 population demography). Also, \citet{Mateos13}, using their IR-selection criterion, show a higher type-1 completeness compared to that observed for type-2 objects at $43\lesssim\log(\rm{L_X[erg\,s^{-1}]})\lesssim44$, while at higher luminosities the type-2 population is likely poorly represented.

In order to explain this trend, \citet{Treister09} invoke the effect of self-absorption based on an analysis using \citet{Nenkova08} clumpy dust-torus models. However, gas may play a role as well. If one considers the scenario where gas is responsible for the bulk of the absorption of X-ray light (see discussion in the following section), and/or hypothesizing type-2 AGN as having intrinsically higher gas obscuration between the X-ray corona and the dust-torus than type-1 AGN, then one should expect weaker IR emission from type-2 AGN (considering the same amount of circumnuclear dust). Although we are unable to confirm what absorber is dominant in our sample, the gas hypothesis cannot be ruled out as an ingredient to produce this bias toward type-1 AGN. Nevertheless, by basing their analysis mostly on optical-spectroscopy, \citet{Treister09} are mostly unaffected by gas absorption (which affects X-rays), thus allowing that dust self-absorption can in principle be behind this type-1-bias in AGN IR-selection.

The general picture on the overall bias observed for each AGN IR-selection criterion is detailed in Table~\ref{tab:typecompG}. It shows a clear bias towards more X-ray luminous sources and, at a lower level, towards type-1 objects. The latter is more noticeable in S50 and KI criteria. D12, KI, and KIM are the criteria that show the highest high-luminosity fraction (i.e., a higher fraction of high-luminosity AGN).

\begin{table*}
\scriptsize
\caption{AGN-type selection comparison.}
\label{tab:typecompG}
\centering
\begin{tabular}{cc|cccccc|cccccc}
\hline\hline
& & \multicolumn{6}{|c|}{CDFS} & \multicolumn{6}{c}{COSMOS} \\
Sample & Criterion & A$_1$\tablefootmark{a} & A$_2$\tablefootmark{a} & $\mathcal{F}_{\rm 2}$\tablefootmark{b} & A$_{\rm H}$\tablefootmark{a} & A$_{\rm L}$\tablefootmark{a} & $\mathcal{F}_{\rm H}$\tablefootmark{b} & A$_1$\tablefootmark{a} & A$_2$\tablefootmark{a} & $\mathcal{F}_{\rm 2}$\tablefootmark{b} & A$_{\rm H}$\tablefootmark{a} & A$_{\rm L}$\tablefootmark{a} & $\mathcal{F}_{\rm H}$\tablefootmark{b} \\
\hline
\smallskip
$K_s$+IRAC & [none]        & 44    & 129   & 0.75~($_{0.03}^{0.03}$)                       & 69    & 149   & 0.32~($_{0.03}^{0.03}$)   & 543   & 515   & 0.49~($_{0.02}^{0.02}$)                       & 961   & 457   & 0.68~($_{0.01}^{0.01}$) \\ 
\smallskip
& L07           & 35    & 78    & 0.69~($_{0.05}^{0.04}$,$_{0.01}^{0.02}$)      & 62    & 83    & 0.43~($_{0.04}^{0.04}$,$_{0.03}^{0.02}$)   & 470   & 376   & 0.44~($_{0.02}^{0.02}$,$_{0.02}^{0.02}$)      & 866   & 245   & 0.78~($_{0.01}^{0.01}$,$_{0.02}^{0.02}$) \\ 
\smallskip
& S05           & 24    & 44    & 0.65~($_{0.06}^{0.06}$,$_{0.09}^{0.09}$)      & 39    & 41    & 0.49~($_{0.05}^{0.05}$,$_{0.13}^{0.09}$)   & 413   & 304   & 0.42~($_{0.02}^{0.02}$,$_{0.04}^{0.03}$)      & 749   & 168   & 0.82~($_{0.01}^{0.01}$,$_{0.03}^{0.04}$) \\ 
\smallskip
& D11           & 14    & 26    & 0.65~($_{0.08}^{0.07}$,$_{0.16}^{0.08}$)      & 24    & 23    & 0.51~($_{0.07}^{0.07}$,$_{0.11}^{0.13}$)   & 277   & 194   & 0.41~($_{0.02}^{0.02}$,$_{0.04}^{0.05}$)      & 514   & 56    & 0.90~($_{0.01}^{0.01}$,$_{0.04}^{0.02}$) \\ 
\smallskip
& KI            & 28    & 55    & 0.66~($_{0.05}^{0.05}$,$_{0.01}^{0.02}$)      & 52    & 48    & 0.52~($_{0.05}^{0.05}$,$_{0.02}^{0.04}$)   & 392   & 289   & 0.42~($_{0.02}^{0.02}$,$_{0.01}^{0.02}$)      & 759   & 118   & 0.87~($_{0.01}^{0.01}$,$_{0.03}^{0.02}$) \\ 
\hline
\smallskip
$K_s$+IRAC+ & [none]        & 35    & 106   & 0.75~($_{0.04}^{0.03}$)                       & 61    & 110   & 0.36~($_{0.04}^{0.04}$)   & 381   & 322   & 0.46~($_{0.02}^{0.02}$)                       & 600   & 238   & 0.72~($_{0.02}^{0.02}$) \\ 
\smallskip
MIPS$_{24\,\mu\rm{m}}$ & L07           & 31    & 61    & 0.66~($_{0.05}^{0.05}$,$_{0.01}^{0.03}$)      & 54    & 62    & 0.47~($_{0.05}^{0.05}$,$_{0.02}^{0.02}$)   & 353   & 259   & 0.42~($_{0.02}^{0.02}$,$_{0.01}^{0.01}$)      & 567   & 156   & 0.78~($_{0.02}^{0.01}$,$_{0.01}^{0.01}$) \\ 
\smallskip
& S05           & 22    & 37    & 0.63~($_{0.07}^{0.06}$,$_{0.06}^{0.08}$)      & 36    & 33    & 0.52~($_{0.06}^{0.06}$,$_{0.11}^{0.08}$)   & 318   & 222   & 0.41~($_{0.02}^{0.02}$,$_{0.03}^{0.01}$)      & 520   & 105   & 0.83~($_{0.02}^{0.02}$,$_{0.02}^{0.02}$) \\ 
\smallskip
& D11           & 12    & 24    & 0.67~($_{0.08}^{0.07}$,$_{0.15}^{0.04}$)      & 22    & 19    & 0.54~($_{0.08}^{0.08}$,$_{0.05}^{0.13}$)   & 234   & 161   & 0.41~($_{0.02}^{0.03}$,$_{0.02}^{0.02}$)      & 406   & 45    & 0.90~($_{0.02}^{0.01}$,$_{0.02}^{0.01}$) \\ 
\smallskip
& KI            & 24    & 43    & 0.64~($_{0.06}^{0.06}$,$_{0.00}^{0.01}$)      & 45    & 35    & 0.56~($_{0.06}^{0.05}$,$_{0.01}^{0.03}$)   & 294   & 206   & 0.41~($_{0.02}^{0.02}$,$_{0.00}^{0.01}$)      & 509   & 75    & 0.87~($_{0.01}^{0.01}$,$_{0.01}^{0.01}$) \\ 
& KIM           & 19    & 37    & 0.66~($_{0.07}^{0.06}$,$_{0.03}^{0.05}$)      & 36    & 32    & 0.53~($_{0.06}^{0.06}$,$_{0.07}^{0.05}$)   & 237   & 170   & 0.42~($_{0.02}^{0.02}$,$_{0.03}^{0.03}$)      & 422   & 59    & 0.88~($_{0.02}^{0.01}$,$_{0.01}^{0.01}$) \\
\hline
\end{tabular}
\tablefoot{
While in the upper group of rows reliable photometry is required --- a magnitude error below 0.36 --- in $K_s$+IRAC bands, in the lower group of rows we also require reliable 24\,${\mu}$m photometry. A$_1$ stands for AGN type-1, whereas A$_2$ for type-2 (X-ray or spectroscopic classifications). A$_{\rm L}$ refers to the sources having intrinsic luminosities of $\log({\rm L_{XR}}[erg\,s^{-1}])<43.5$, while A$_{\rm H}$ refers to those having $\log({\rm L_{XR}}[erg\,s^{-1}])\geq43.5$.\\
\tablefoottext{a}{Number of AGN sources selected by the applied IR criterion.}\\
\tablefoottext{b}{AGN type fraction: $\mathcal{F}_{\rm 2}=\rm{A_2/(A_1+A_2)}$ and $\mathcal{F}_{\rm H}=\rm{A_H/(A_H+A_L)}$. Higher $\mathcal{F}_{\rm 2}$ or $\mathcal{F}_{\rm H}$ values than those found for the whole sample (first row in each row group) imply a bias toward A$_2$ or A$_{\rm H}$ AGN, respectively. Numbers in parenthesis refer, respectively, to the statistical error and the photometric upper and lower error bars.}\\
\tablefoottext{c}{The first row in each group refers to the total number of sources of a given type with reliable $K_s$+IRAC (upper group) and $K_s$+IRAC+24$\,\mu$m (bottom group) photometry.}
}
\end{table*}

\section{Discussion on AGN IR-selection completeness} \label{sec:ircomp}

In Paper I, it was shown an overall low completeness ($\lesssim50\%$) for the CDFS sample, especially for S05, D12, KI, and KIM criteria at $z<2.5$. At higher redshifts, or brighter flux limits (as in the COSMOS sample), one is restricted to more luminous objects, which are prone to be selected by IR AGN diagnostics \citep[][and Sect.~\ref{sec:selecLum}]{Treister06,Cardamone08,Donley08,Eckart10}, resulting in high completeness levels. With decreasing redshifts, however, X-rays observations and optical spectroscopy are able to detect more low-luminosity AGN, sources which will likely not dominate the total IR SED of a galaxy either resulting from IR-luminous obscured star-formation in the galaxy \citep{Rigopoulou99,Veilleux09,Nordon11} or from the complete absence of a dust torus \citep[e.g.,][]{Ho08,Asmus11,Burlon11,Plotkin11}. The former plays a significant role, as IR-selected AGN have been mostly found in star-forming hosts \citep{Hickox09,Griffith10}.

But one key reason, which affects the completeness at all redshifts, is the fact that the major source of obscuration of AGN light is gas. Short time variability for both flux and absorption column density \citep{Elvis04,Risaliti07,Giustini11} implies obscuration material at smaller radii than the dusty torus inner radius, which is set by the sublimation radius \citep[R$_{sub}$, e.g.,][]{Suganuma06,Nenkova08}. These dust-free gas clouds will only absorb X-ray emission, and will not re-emit a continuum spectrum in the IR as dust does. Also, the dusty material, which absorbs UV/optical light, blocks both X-ray and UV/optical emission. Hence, X-ray column densities are always larger than those producing the UV/optical obscuration, up to extreme ratios of two orders of magnitude \citep{Maccacaro82,Gaskell07}. These findings imply that dust-free clouds at $<\rm{R_{sub}}$ frequently represent the bulk of the X-ray obscuration. As such, the AGN-induced dust emission will be relatively weak or non-existent especially in low-luminosity AGN \citep{Ho08,Asmus11,Plotkin11}, but high-luminosity cases also exist where the IR excess is not as extreme as expected \citep{Comastri11}.

The observed low completeness levels \emph{should not} be taken as the real completeness levels of AGN IR-selection criteria. At this stage it is impossible to estimate the incompleteness of the X-rays and optical criteria used to assemble our base control sample (Sect.~\ref{sec:samp}). This result just means that, at lower redshifts there is a population of low-luminosity AGN which will be successfully detected in the X-rays and optical regimes, but completely outshone by the host galaxy light in the IR regime.

\section{Strategy on AGN IR-selection} \label{sec:strat}

With such a plethora of AGN IR-selection criteria, the astronomer faces a hard decision on which one to adopt. As evidenced by the many techniques and criteria existent in the literature, different methods will either select a heterogeneous AGN sample or a sub-set of it, and with different efficiencies. Based on the results from Paper I and this work, as well as those found in the literature, we propose different strategies depending on survey/sample characteristics to assess the IR AGN population.

If a spectroscopic redshift is available, the astronomer is advised to use:
\begin{itemize}
\item[-] the $K_s-[4.5]$ colour or the \citet{Assef13} criterion\footnote{$\rm{W1-W2>0.662~exp\left\{0.232 (W2-13.97)^2\right\}}$, where W1 and W2 are, respectively, the 3.4$\,\mu$m and 4.6$\,\mu$m bands on board WISE.} at $z<1$;
\item[-] the $[4.5]-[8.0]$ colour at $1<z<2.5$;
\item[-] the $[8.0]-[24]$ at $z>3$.
\end{itemize}

On the other hand, if a spectroscopic redshift is not available and if a proper photometric redshift has not been estimated \citep[as in, e.g.,][where photometric redshift where tuned for AGN emission]{Salvato09, Salvato11, Cardamone10, Luo10, Fotopoulou12}, the astronomer is advised to use:
\begin{itemize}
\item[-] the \citet{Assef13} criterion or the $K_s-[4.5]$ colour aided by a flux cut \emph{a la} \citet{Assef13} in samples where only WISE data is available and/or no deep $\gtrsim5\,\mu$m data is available;
\item[-] a criterion composed by $<8\,\mu$m spectral-bands in shallow/intermediate-depth coverages, such as L07, S05, D12, or KI, where KI is believed to be the best compromise between completeness and reliability;
\item[-] the KIM criterion for deep coverages or a criterion composed by $<8\,\mu$m spectral-bands aided by a flux cut \emph{a la} \citet{Assef13} if no deep $24\,\mu$m data is available;
\item[-] extreme optical-to-IR flux ratios (e.g., $f_{24}/f_R>1000$) in addition to bright flux cuts and/or additional colour cuts \citep[e.g.,][]{Fiore08,Polletta08}, in case the goal is the selection of rare extremely obscured AGN alone.
\end{itemize}

\section{Adapting KIM to the JWST filter set} \label{sec:jwst}

\begin{figure}
\centering
\includegraphics[width=\hsize]{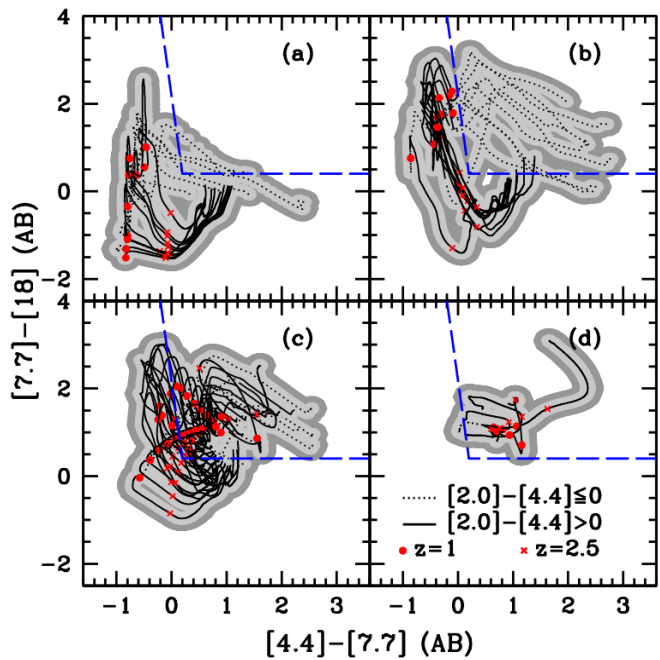}
\caption{Model color tracks displayed in the proposed color-color space with \textit{JWST} bands adapted from KIM criterion. Each panel presents a specific group: (a) Early/Late, (b) starburst, (c) Hybrid and (d) AGN. The dotted portion of the tracks refers to the redshift range where $[2.0]-[4.4]\leq0$, and $[2.0]-[4.4]>0$ when solid. Red circles and stars along the lines mark $z=1$ and $z=2.5$, respectively. The light and dark-grey regions show the photometric scatter due to a magnitude error of, respectively, 0.1 and 0.2 in the considered bands (equivalent to $\sim10\%$ and $\sim20\%$ error in flux, respectively).}
\label{fig:jwstnew}
\end{figure}

The adaptation of KIM to planned \emph{JWST} filters is not straightforward. Near-to-mid-IR filters used in KIM do have different responses from those of \emph{JWST}. Also, the \emph{JWST} filter set is richer. Preferably, the medium-band $K_s$ is replaced by the 2.0$\,\mu$m broad-band \emph{JWST} filter, allowing deeper flux levels to be reached. As for the \emph{Spitzer}-IRAC 4.5$\,\mu$m filter, the \emph{JWST} 4.4$\,\mu$m filter is the alternative. As for the 8$\,\mu$m and 24$\,\mu$m filters on board \emph{Spitzer}, in Paper I, we have proposed the 10$\,\mu$m and 21$\,\mu$m filters as preliminary alternatives. However, using the 10$\,\mu$m filter creates a degeneracy between high-redshift AGN and $z\sim1$ highly obscured non-AGN starbursts in the 2--8\,$\mu$m colour space. Although a morphology analysis could in principle correct this, we now propose filters 7.7$\,\mu$m and 18$\,\mu$m as the new alternatives for KIM filters 8$\,\mu$m and 24$\,\mu$m (Fig.~\ref{fig:jwstnew}). Besides preventing the $z\sim1$ non-AGN contamination, these two filters are less affected by telescope thermal emission, thus allowing fainter exposures. The new colour constraints are the following:
\begin{equation}
  \rm{KIM_{JWST}} \equiv 
  \begin{cases}
    \left[2.0\right]-\left[4.4\right]>0 \\
    \left[7.7\right]-\left[18\right] > 0.4 \\
    \left[7.7\right]-\left[18\right] > -9\times(\left[4.4\right]-\left[7.7\right])+2.2
  \end{cases}
\end{equation}

The possible disadvantage, is the slight contamination by highly obscured non-AGN galaxies at $z>6$. Their existence, however, is at this stage unknown. Although $z>6$ galaxies have been found already, they show low levels of metallicity \citep[$\lesssim\rm{Z}_\odot$,][]{Dunlop12,Dunlop13,Zitrin12,Jiang13}, consequently, low or no dust content. These will appear much bluer in both $[4.4]-[7.7]$ and $[7.7]-[18]$ colours. But the rest-frame UV selection in these works, on the other hand, may be inducing a selection bias missing a dust-rich population. For instance, gravitational lensing has recently revealed two dusty sources at $z\sim5.7$ \citep{Vieira13}, and large amounts of dust have indeed been found at $z>6$, even though only in a QSO host-galaxy so far \citep{Bertoldi03,Gall11,Valiante11}.

\section{Conclusions} \label{sec:conc}

This work tests the dependency of AGN selection via IR-colour techniques on source luminosity and obscuration. For that purpose we have assembled a deep IR AGN sample in the CDFS and COSMOS fields, and characterised it by X-ray luminosity, and by obscuration as indicated by X-rays or optical spectroscopy.

We observe the already known strong bias toward high-luminosity AGN, and show that the bias toward type-1 (unobscured) sources is only significant at intermediate luminosities ($43.3\lesssim\log(\rm{L_X[erg\,s^{-1}]})\lesssim44$). At higher luminosities, IR criteria select with equal efficiency type-1 and type-2 AGN, while at lower luminosities, IR criteria have large incompleteness levels, thus increasing dramatically the statistical error. Hence, AGN IR-selection is genuinely biased toward unobscured AGN, but only at intermediate luminosities. However, we show that type-1 AGN are intrinsically more luminous than type-2 AGN at $z\lesssim1.6$, thus resulting in more type-1 AGN being selected the shallower the IR survey is, as a result from the IR-selection dependency on luminosity (Sec. \ref{sec:typesel}).

We explicitly show the reader that survey characteristics and method used to assess AGN obscuration do matter. Disregarding this fact will induce large discrepancies in the comparison of results (Sec.~\ref{sec:t12dem}).

These results strengthen the strategy where (i) to gather a complete or more reliable AGN sample, one should consider a multi-wavelength set of AGN criteria, (ii) deep X-ray observations are important to recover the low-luminosity AGN, and (iii) hard-X-ray observations (such as those with NuSTAR), and/or IR-radio selection, and/or high-resolution IR imaging (e.g., with \emph{JWST}) are key to recover the (highly-)obscured AGN, especially at low-luminosities.

Specifically for the AGN IR colour-selection, and considering our study in Paper I and here, as well as work in the literature, we propose a selection strategy dependent on the availability of a reliable redshift measurement, or survey depth, or target AGN population (Sec.~\ref{sec:strat}).

Finally, we point out that when \emph{JWST} comes online, the broad-band filters 2.0$\,\mu$m, 4.4$\,\mu$m, 7.7$\,\mu$m, and 18$\,\mu$m will be key to disentangle AGN from non-AGN dominated SEDs at depths where spectroscopy becomes impractical, or for AGN selection in case a \emph{JWST} deep survey is pursued (Sec. \ref{sec:jwst}).

\begin{acknowledgements}
The authors thank the anonymous referee who carefully read the manuscript and helped improving it. The authors thank the MUSIC, COSMOS, \citeauthor{Xue11} groups for providing public catalogues with key information, which made possible the present work. HM acknowledges the frequent use of Topcat and VOdesk. HM acknowledges the support from Funda\c{c}\~{a}o para a Ci\^{e}ncia e a Tecnologia through the scholarship SFRH/BD/31338/2006 and the support by CONICyT-ALMA through a post-doc scholarship under the project 31100008. HM and JA acknowledge support from Funda\c{c}\~{a}o para a Ci\^{e}ncia e a Tecnologia through the projects PTDC/CTE-AST/105287/2008 and PEst-OE/FIS/UI2751/2011. HM acknowledges the support by UCR while visiting Dr. Bahram Mobasher as a visitor scholar. MS acknowledges support by the German Deutsche Forschungsgemeinschaft, DFG Leibniz Prize (FKZ HA 1850/28-1).
\end{acknowledgements}

\appendix
\section{Biases affecting type-1/type-2 demography} \label{sec:t12dem}

Besides a possible natural luminosity-obscuration relation and its evolution with redshift (Sections~\ref{sec:selecLum} and \ref{sec:z12}), assessing the obscuration level of an AGN is dependent on the technique used for that purpose, thus influencing the final conclusions. Here, we explicitly show the differences.

X-ray obscuration can be assessed via spectroscopic or photometric analysis. While the former is the most reliable, it also limits the analysis to the brightest objects. Hence, the Hardness-Ratio (HR) or the column density ($\rm{N_H}$) are estimated photometrically and commonly adopted. The HR is based on photon count ratio (i.e., (H-S)/(H+S), where H and S are the hard- and soft-band counts), which depends on telescope soft-to-hard band relative sensitivity, and is biased toward type-1 objects with increasing redshift \citep{Eckart06,Messias10}. As described in Paper I, $\rm{N_H}$ is estimated based on soft-to-total flux-ratio and redshift, but it requires a correction due to a bias toward type-2 objects with increasing redshift and flux error \citep[Paper I, but see also][]{Akylas06,Donley12}.

In Tab.~\ref{tab:agnfrac}, the different approaches are compared considering the CDFS and COSMOS separately and together. The first row shows the type-2 fractions when HR and the obscured luminosity are adopted. This is the method which retrieves the lowest values. Adopting $\rm{N_H}$ (second row), the increase is significant, especially in COSMOS where it doubles. If one then corrects the observed luminosity to an intrinsic luminosity (third row), the slight differences compared to the second row result from changes in the low-luminosity population (mostly due to the different luminosity lower-limit in the type-1 and type-2 X-ray classification). The bottom row shows the result of correcting the $\rm{N_H}$ estimate, which yields no significant difference in COSMOS, but a significant decrease in CDFS compared to the third row. This is mostly due to the higher fraction of high-redshift sources in the CDFS sample (Fig.~\ref{fig:redist}) for which the $\rm{N_H}$ correction is most relevant.

\begin{table}
\caption{Assessing the obscured fraction through different methods.}
\label{tab:agnfrac}
\centering
\begin{tabular}{crrr}
\hline\hline
Parameters & CDFS & COSMOS & Overall \\
\hline
\smallskip
$\rm{L_X^{OBS}+HR}$ & $0.72^{0.03}_{0.04}$ & $0.21^{0.01}_{0.01}$ & $0.28^{0.01}_{0.01}$ \\
\smallskip
$\rm{L_X^{OBS}+N_H}$ & $0.85^{0.03}_{0.03}$ & $0.44^{0.02}_{0.02}$ & $0.51^{0.02}_{0.02}$ \\
\smallskip
$\rm{L_X^{INT}+N_H}$ & $0.84^{0.03}_{0.03}$ & $0.48^{0.02}_{0.02}$ & $0.53^{0.01}_{0.01}$ \\
$\rm{(L_X^{INT}+N_H)}$\tablefootmark{a} & $0.73^{0.03}_{0.04}$ & $0.47^{0.02}_{0.02}$ & $0.51^{0.01}_{0.01}$ \\
\hline
\end{tabular}
\tablefoot{
The listed values are the type-2 AGN fraction, $\rm{A_2/(A_1+A_2)}$,  where $\rm{A_1}$ and $\rm{A_2}$ are the numbers of type-1 and type-2 sources, respectively. $\rm{L_X^{OBS}}$ and $\rm{L_X^{INT}}$ are, respectively, the observed and intrinsic (obscuration corrected) luminosities, while HR and N$_{\rm H}$ are, respectively, the Hardness-Ratio and column density (two different alternatives to assess the AGN type-1 and type-2 populations).\\
\tablefoottext{a}{Values have been statistically corrected for the bias affecting the N$_{\rm H}$ estimate via flux ratios (see Paper I).}
}
\end{table}

\begin{figure*}[t]
\centering
\includegraphics[width=\hsize]{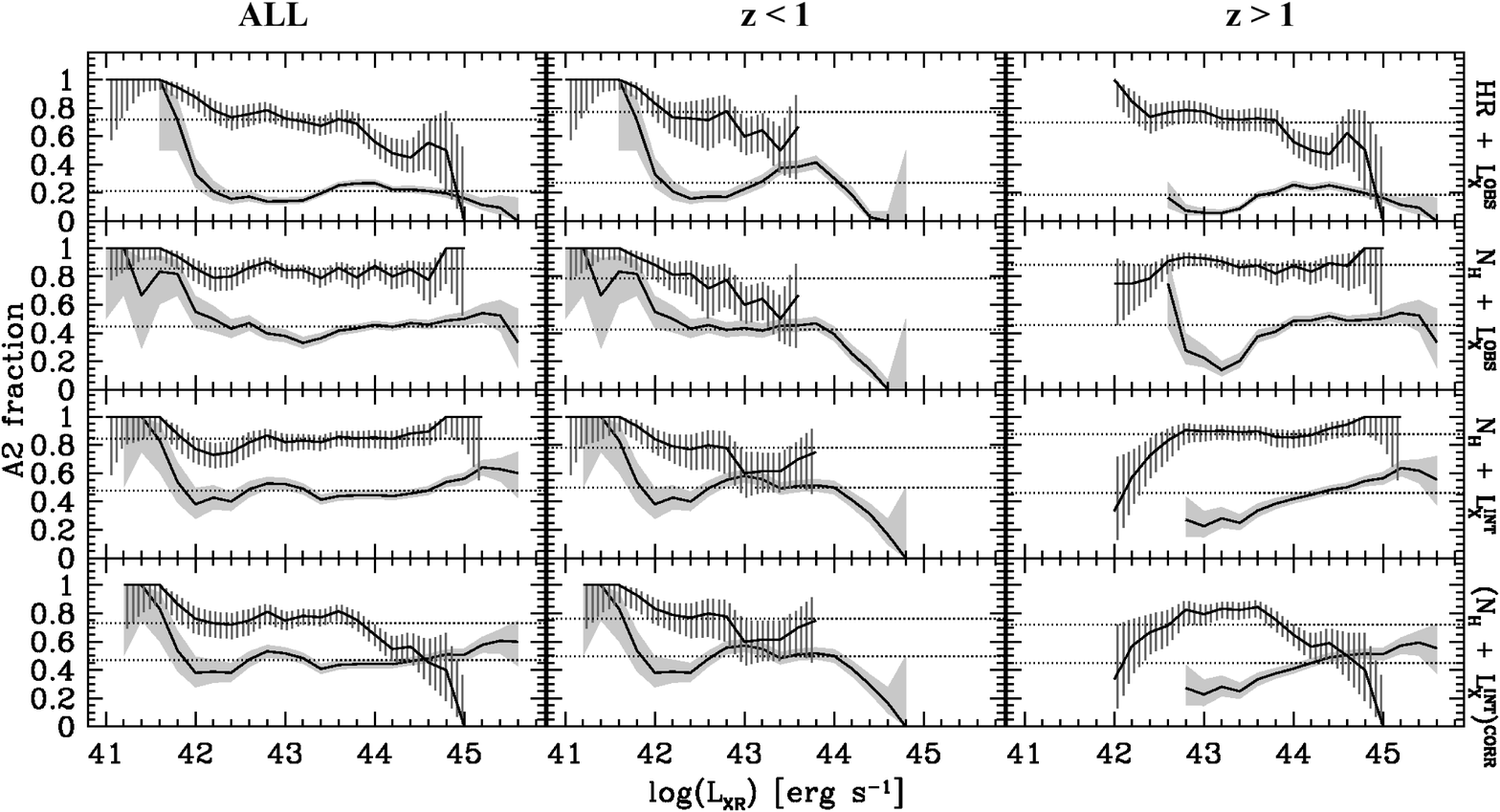}
\caption{The variation of the $\rm{A_2}$ fraction with source X-ray luminosity in CDFS (solid line with hatched uncertainty region) and COSMOS (solid line with shaded uncertainty region). The ratio is $\rm{A_2}~/~(\rm{A_1}+\rm{A_2})$, where $\rm{A_1}$ and $\rm{A_2}$ are the numbers of type-1 and type-2 sources, respectively. Here, the two samples are considered individually and at two different redshift ranges (middle and right columns). The different panels show the effect of different assumptions while assessing the luminosity and obscuration classes. The two upper panels show the difference between considering HR or $\rm{N_H}$ to separate the population into type-1 (unobscured) and type-2 (obscured) sources. The middle lower panel shows the effect of adopting the intrinsic luminosity ($\rm{L_X^{INT}}$) instead of the observed luminosity ($\rm{L_X^{OBS}}$, upper panels). Finally, the bottom panel shows the effect of correcting for the $\rm{N_H}$ bias when derived via flux-ratios (Paper I). A moving bin is used as described in Fig.~\ref{fig:lxdist}. The horizontal dotted line shows the overall $\rm{A_2}$ fraction in each case, $\rm{A_2^{TOT} /~( A_1^{TOT} + A_2^{TOT} )}$.}
\label{fig:p21tech}
\end{figure*}

The left column in Fig.~\ref{fig:p21tech} shows how the values reported in Tab.~\ref{tab:agnfrac} depend on luminosity. In CDFS the differences come mostly from discrepancies at $\log({\rm L_{XR}[erg\,s^{-1}}])\gtrsim43.6$, while in COSMOS it is an overall shift at $\log({\rm L_{XR}[erg\,s^{-1}}])\gtrsim42$. The middle column shows the same, but only for AGN at $z<1$. While the CDFS sample does not show significant differences between methods, the COSMOS samples does show a clear difference between adopting the HR or $\rm{N_H}$ (see the two top panels) at $\log({\rm L_{XR}[erg\,s^{-1}}])\lesssim43.4$ (range where COSMOS is likely incomplete, Fig.~\ref{fig:s21lumx}). This can be explained either by the COSMOS survey depth affecting more the HR analysis than the $\rm{N_H}$ analysis of the less-luminous objects, or this behaviour is being missed in CDFS given the sample's higher statistical uncertainty and/or small volume. At the highest redshifts (right hand-side column), the COSMOS sample shows again an overall shift to higher type-2 fractions when adopting $\rm{N_H}$ instead of HR, while the CDFS sample shows major differences at $\log({\rm L_{XR}[erg\,s^{-1}}])\gtrsim43.6$ when statistically correcting the $\rm{N_H}$ method. One possible reason explaining why one does not observe the same behaviour at these luminosities in the COSMOS sample is that, although both samples are limited to $z>1$, the COSMOS statistics are still dominated by the $z\sim1-2$ objects (62\% of the $z>1$ COSMOS sample with $\log({\rm L_{XR}[erg\,s^{-1}}])>43.6$), distances at which the correction is still not considerable. However, the $z>1$ CDFS sample has a higher fraction of $z>2$ (65\% of the $z>1$ CDFS sample with $\log({\rm L_{XR}[erg\,s^{-1}}])>43.6$) where the correction starts to be necessary.

Nevertheless, based on the method we have adopted (results shown in the bottom panels), it is interesting to see that within the uncertainties the CDFS trends tend to the values of the COSMOS sample at high-luminosities, where the COSMOS survey is expected to be complete.

\end{document}